\documentclass[fleqn]{2017SCGE}
\usepackage{graphicx}
\newcommand\cL{\mathcal{L}}
\begin{document}

\ensubject{Theoretical Physics}

\ArticleType{Article}
\SpecialTopic{}
\Year{2018}
\Month{December}
\Vol{61}
\No{12}
\DOI{10.1007/s11433-018-9251-0}
\ArtNo{000000}
\ReceiveDate{April 6, 2018}
\AcceptDate{May 22, 2018}
\OnlineDate{June 15, 2018}

\title{The CST Bounce Universe model
-- A Parametric Study}{The CST Bounce Universe model
-- A Parametric Study}

\author[1,2]{Yeuk-Kwan E. Cheung}{cheung@nju.edu.cn}
\author[1,3]{Xue Song}{}
\author[1,4]{ShuYi Li}{}
\author[1,5]{YunXuan Li}{}
\author[1,6]{YiQing Zhu}{}

\AuthorCitation{Y.-K. E. Cheung, X. Song, S. Y. Li, Y. X. Li, and
Y. Q. Zhu}

\address[1]{Department of Physics, Nanjing University, Nanjing 210098, China}
\address[2]{Institute of High Energy Physics, Beijing, China}
\address[3]{Department of Physics, Princeton University,
Princeton NJ  08544, USA}
\address[4]{Department of Physics, Brown University,
Providence RI 02912, USA}
\address[5]{California Insitute of Technology, Pasadena  CA 91125, USA}
\address[6]{ETH Z\"{u}rich, 8092
Z\"{u}rich, Switzerland}

\abstract{A bounce universe model with a scale-invariant and stable spectrum of primordial density perturbations was constructed using a consistent truncation of the D-brane dynamics from Type IIB string theory. A coupling was introduced between the tachyon field and  the adjoint Higgs field on the D3-branes  to lock the tachyon at the top of its potential hill  and to  model the bounce process, which is known as the Coupled Scalar and Tachyon Bounce (CSTB) Universe. The CSTB model has been shown to be ghost free, and it fulfils the null energy condition; in addition, it can also  solve the Big Bang cosmic singularity problem. In this paper we conduct an extensive follow-up study of the parameter space of the CSTB model. In particular we are interested in the parameter values that can produce a single bounce to arrive at a  radiation-dominated universe. We further establish  that the CSTB universe is a viable alternative to inflation, as  it  can naturally  produce a sufficient number of e-foldings in the locked inflation epoch and in the post-bounce expansion to overcome the four fundamental limitations of the Big Bang cosmology, which are flatness, horizon, homogeneity and singularity, resulting in a universe of the current size.}



\maketitle



\begin{multicols}{2}
\section{Introduction}
\label{sec:intro}

The four  most fundamental problems in the Big Bang cosmology are Flatness,
Horizon, Homogeneity and Singularity.
In the 1980s inflation was proposed~\cite{Inflation} (with an update in~\cite{update})  to address the first three of these problems.
The predictions of the inflation paradigm are consistent with the
high-precision
cosmological observations,  such as the the highly isotropic and homogeneous cosmic microwave background radiation (CMBR)~\cite{CMB}, 
and the scale invariance of primordial density perturbations attributed 
to the quantum fluctuations of the inflaton field at the end of inflation~\cite{Mukhanov:1990me}.
Inflation, however, has its problems. For instance, the initial singularity does not disappear but is  delayed~\cite{Borde:1993xh}.

Building on the success of inflation models,
cosmologists have been searching for alternative scenarios
that can  address the problem of the cosmic singularity.
In 1993 a GR-compatible solution with  a nonsingular bounce was proposed
shortly after the first observation of current
expansion~\cite{93Solve}.
Meanwhile string theorists attempted to implement a nonsingular bounce in dilaton gravity, leading to the pre-big-bang (PBB) scenario within the string
theory framework~\cite{PBB}.  A period of intensive string-inflation model building ensues, as reviewed by Linde~\cite{Linde:2009zz}, and
Tye~\cite{HenryTye:2006uv}.

The bounce universe itself has a long history,
and many ingenious models
are reviewed in~\cite{reviews}.
 A cyclic model in a conventional
four-dimensional quantum field theory generates a bounce that  solves the singularity problem~\cite{Steinhardt:2001vw}.
``Matter bounce'' postulates a period of contraction, in which the scale invariant
 spectrum  of primordial  density perturbations is
generated~\cite{Wands:1998yp, Finelli:2002prd},
prior to the Big Bounce; at the Big Bounce
 the universe turns around and enters  a period of expansion resulting
  in our existing  universe.
Explicit model building follows.
Ghost condensation was proposed to avoid ghosts in the
perturbation spectrum~\cite{
matterbounce, Brandenberger:2009yt, Cai:2011tc, Odintsov:2014gea, Qiu:2010ch}.
 \textit{N}=1 supergravity is used in a bounce model to eliminate  ghost excitations~\cite{N=1} and to prove that nonsingular bounces were viable in supergravity.
  ``G-Bounce'' models  generate a non-singularity inflation cosmos with a Galileon field~\cite{Gbounce}, while Lee-Wick bounce produces  a bounce in the radiation phase with an apparent scale invariance in the  perturbation
  spectrum~\cite{Cai:2008qw, LWradiation}.
A bounce solution is  also found in the universe dominated by the Quintom matter~\cite{Quintom1, Quintom2}. Furthermore the well-known  Ekpyrotic
universe~\cite{Ekpyrotic} is inspired by D-branes dynamics.
Bounce conditions were also found possible in $f(R)$ cosmology~\cite{Carloni:2005ii}.

Encouraged by these advances in bounce cosmology  we  improved on the original
 tachyon inflation model proposed independently by 
 Sen~\cite{Sen:2002nu, Sen:2002in} and  Gibbons~\cite{Gibbons:2002md}, 
 by  incorporating an interaction between
 the tachyon fields and adjoint Higgs fields.
 Thus we can  not only  we greatly improve the number of e-foldings obtained from the model by locking
  the tachyon at the top of its potential hill with the Higgs fields,
  but  also realize a bounce universe in string theory  with this consistent truncation
  of the D-brane dynamics.
  We call this the  coupled scalar-tachyon
  bounce universe model or CSTB for short~\cite{CSTB, Li:2012vi, Li:2013bha}.

 The CSTB model has several  salient features,  such as the generation of
  stable and scale invariant primordial density  perturbations,
 absence of ghosts,  and no violation of the null energy condition throughout
 the  cosmic evolution~\cite{CSTB}.
 It thus motivates  us to investigate bounce universe models further  and
 derive  testable predictions in a model-independent manner.
 Dark matter creation and evolution is studied in the bounce
 universe~\cite{Li:2014era, Cheung:2014nxi, Cheung:2014pea, Cheung:2016wik, Vergados:2016niz}. When sufficient dark matter particles are
 not produced to reach kinematic equilibrium during contraction,
 relic abundance  in the present universe then imposes  a constraint on the
 dark matter mass and coupling constant.
 We can then produce numerical predictions for heavy  dark matter
  in nuclear recoils experiments~\cite{Cheung:2014pea}  and for
   light dark matter in electron-scattering experiments~\cite{Vergados:2016niz}.

In the rest  of this paper,  after providing a short review of the CSTB model,
 we present   a thorough study of the parameter space of this model.
 Our results  verify   that CSTB can produce a sufficient number of
  e-foldings to
 solve the problems related to flatness, horizon,  homogeneity and Big Bang singularity, which will result in
  a universe of the current size.
Furthermore
  this is achieved with a large volume of the parameter space with no limitation
  on the values of the free parameters in the model.
  In other words the
  problems related to the Big Bang, including the Big Bang problem, are solved naturally in the
  CSTB model.

\section{The Single Bounce Criteria}

In this section, we review briefly the  CSTB model and discuss the essential
physics of this string cosmology model based on D-brane and
$\mathrm{\bar{D}}$-brane  annihilations.
In the effective theory, a D-$\mathrm{\bar{D}}$3-brane pair is described by
the open string tachyon field action~\cite{Sen:2002in},
\begin{eqnarray}
\label{singleaction}
\cL_T=V_{T}\sqrt{1+M_s^{-4}\partial_{\mu}T\partial^{\mu}T},
~ V_{T}=\frac{V_0}{\cosh(\frac{T}{\sqrt{2}M_s})}~~
\end{eqnarray}
where $T$, $M_s$, $V_0$ are the tachyon field, the string mass, and
the tension of D-$\mathrm{\bar{D}}$3-brane pair, respectively.
The tachyon potential $V_{T}$ has a maximum at $T=0$ and two minima at $T\rightarrow\pm\infty$.
We take the background metric as $(-,+,+,+)$, and we also assume that the tachyon field is spatially homogenous.

In D-brane inflation~\cite{braneinflation} the attractive potential of the D-$\bar{D}$ pair takes the following form:
\begin{eqnarray}\label{scalarpotential}
V_{\phi}=\frac{1}{2}m_{\phi}^2\phi^2+V_0-\frac{V_0^2}{4\pi^2\upsilon\phi^4},\qquad \phi\equiv\sqrt{V_0}\,y,
\end{eqnarray}
where $y$ being the distance between the D-$\mathrm{\bar{D}}$3-branes.

The CSTB model~\cite{CSTB, Li:2012vi, Li:2013bha} introduces a scalar-tachyon coupling term
$\lambda\phi^2T^2$, and the tachyon condensation  comes  about naturally after a period of the locked inflation, driven by the tension of D-$\mathrm{\bar{D}}$3-branes, and another period of rolling inflation while rolling down its potential hill.
The coupling of between the scalar and the tachyon plays a crucial role in
the cosmic evolution by  greatly extending the time the tachyon spends at
the top of its potential  hill and hence increasing the number of the e-foldings
obtained in the periods of locked inflation and rolling inflation. This is
other impossible~\cite{Sen:2002nu, Sen:2002in, Gibbons:2002md}.

The CSTB Lagrangian takes the form:
\begin{equation}   \label{action}
\cL= 
\cL_{T}
-\frac{1}{2}\partial_{\mu}\phi\partial^{\mu}\phi-\frac{1}{2}m_{\phi}^2\phi^2-\lambda\phi^2T^2,
\end{equation}
in addition to the 
Hilbert-Einstein term, in a FRW background with $k=1$,
\begin{equation}   \label{metric}
{ds}^2=-{dt}^2 + a^2(t) [\frac{{dr}^2}{1-kr^2}+r^2
(d\theta^2+\sin^2\theta{d\phi}^2)]~.
\end{equation}
Since each component of this model has the correct sign of kinetic energy term and positive potential, we have $\rho_{tot}>0$ and
$\omega_{tot}=\frac{\rho_{tot}}{p_{tot}}\ge-1$,
where $\omega_{tot}$, $\rho_{tot}$ and $p_{tot}$ being, respectively,
the  Equation of State, energy density and pressure derived from the model.
The CSTB  universe model is thus  free of ghosts and satisfies the null energy condition $p+\rho\ge 0$  throughout  the cosmic evolution~\cite{MolinaParis:1998tx}.

\subsection{Cosmic evolution in the CSTB model}
\begin{figure}[htbp]
\centering
\includegraphics[width=2in]{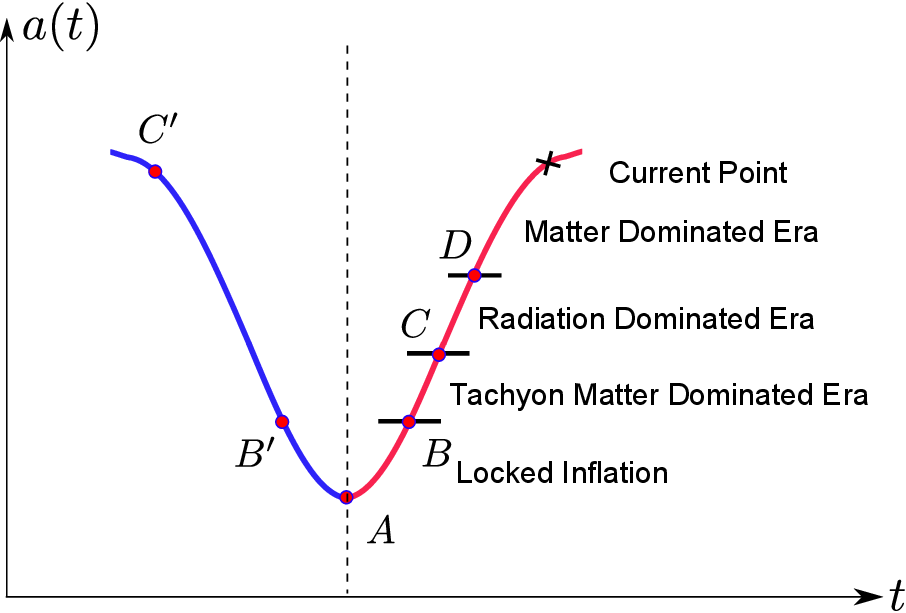}
\caption{The evolution of cosmos in the CSTB model.
Our model starts at  $C^\prime$ as it undergoes a contraction phase,
from $C^\prime$ to $B^\prime$,
 and then a deflation era,  from $B^\prime$ to $A$,
 before entering  a locked inflation era from $A$ to $B$.
 About the bounce point, the exponential expansion of the universe starts, $\dot{a}=0$ and $\ddot{a}>0$. This  period is extended by the effective coupling of
 tachyon and Higgs fields. Quantum mechanically the tachyon is prevented from
 decaying (tachyon condensation) by the fast oscillations of the Higgs fields.
 The tachyon condensation eventually  takes place at $B$, where all the energy of tachyon field transfers to tachyon matter.
 The phase transition at $B$ is followed by a tachyon matter dominated period, $B{\rightarrow}C$.
 At point $C^\prime$, reheating may happen as the  universe undergoes another
 phase transition, during which
 the energy of tachyon matter transfers to radiation.}
 \label{fig:bounce-model}
\end{figure}

The CSTB cosmos  starts at action at $C^\prime$ as it undergoes a period of
tachyon matter dominated contraction, from $C^\prime$ to $B^\prime$,
 during which the tachyon behaves like
cold dark matter~\cite{Sen:2002in, tachyoncondensation},
From  $B^\prime$ to $A$, it undergoes a period of accelerating contraction
as the potential energy comes into play.
 A locked inflation era takes place from $A$ to $B$ (about the bounce point
 $\dot{a}=0$ and $\ddot{a}>0$) as the universe undergoes a
 period of exponential expansion driven by the vacuum energy of the tachyon
 (due to the brane tension).
 This  period is extended by the effective coupling of
 tachyon and Higgs fields, and it solves the problems with tachyon inflation pointed out by~\cite{Kofman:2002rh}. Quantum mechanically the tachyon is prevented from
 decaying (tachyon condensation) by the fast oscillations of the Higgs fields.
 The tachyon condensation eventually  takes place at $B$, where all the energy of tachyon field transfers to tachyon 
 matter~\cite{Sen:2002in, tachyoncondensation}.
 The phase transition at $B$ is followed by a tachyon matter dominated period, $B{\rightarrow}C$.
 At point $C^\prime$, reheating may happen as the  universe undergoes another
 phase transition, during which
 the energy of tachyon matter transfers to radiation.
 This is depicted schematically in Fig.\ref{fig:bounce-model}.

The background cosmology evolves according to the Friedman equations,
\begin{equation}\label{eomH}
H^2 = -\frac{1}{a^2}+\frac{8\pi}{3M_p^2}
\left[
   \frac{V(T)}{\sqrt{1-M_s^{-4}\dot{T}^2}}+
   (\frac{1}{2}m_{\phi}^2+{\lambda}T^2)
   \phi^2+\frac{1}{2}\dot{\phi}^2
\right]; \nonumber
\end{equation}
whereas  the equations of motion for $T$  and $\phi$ are governed by
\begin{eqnarray}
\label{eomT}
\ddot{T}^2
+ (1- 
\dot{T}^2) \left[
3H\dot{T}
+  \frac{V'(T)+   2 \lambda \phi^2 T
\sqrt{1-   
  \dot{T}^2}}{V(T)} \right]
=0,\nonumber \\
%
\label{eomphi}
\ddot{\phi}+3H\dot{\phi}+ (m_{\phi}^2+2{\lambda}T^2)\phi=0; ~~~
\nonumber
\end{eqnarray}
with  the string mass, $M_{s}$, being  suppressed in the above equations.

\begin{figure}[htp!]
\includegraphics[width=2in]{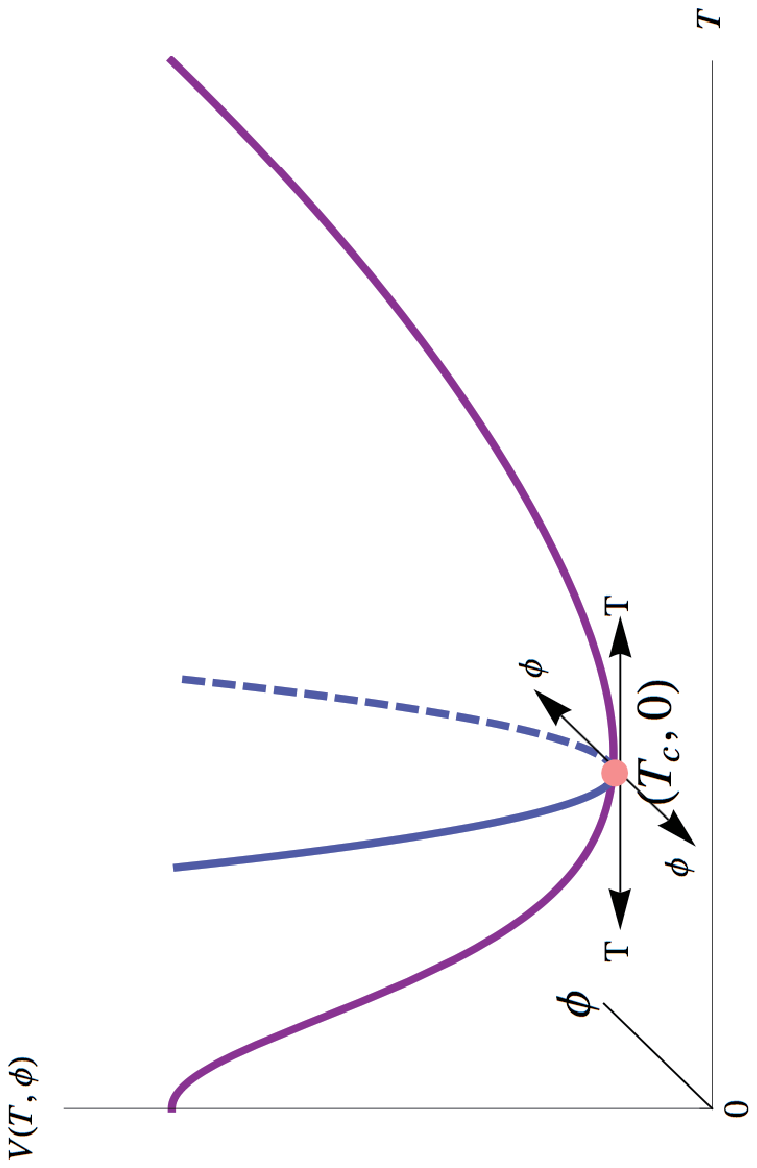}
\caption{A sketch of the effective potential of the CSTB cosmos.
For a non-vanishing value of $\phi$, the coupling term pulls back the vacuum to  finite, $(T_\textmd{c}, 0)$~. The tachyon is, therefore, stabilised around the vacuum with quasi-harmonic oscillations. 
This plot is from~\cite{Li:2012vi, Li:2013bha}.}
\label{fig:Potential}
\end{figure}

The solutions to  these equations of motion have been studied~\cite{CSTB} in detail.
It turns out, by introducing the coupling term $\lambda T^2\phi^2$,  CSTB model
suggests a novel feature: for a non-vanishing value of $\phi$, the coupling term pulls back  the vacuum expectation value of this model at finite, $(T, \phi)=(T_\textmd{c}, 0)$, as shown in 
Figure~\ref{fig:Potential} and in~\cite{Li:2012vi, Li:2013bha}.
In contrast to the single tachyon field model which tachyon keep rolling towards infinity~\cite{Sen:2002nu}, the tachyon in this model is stabilised around the effective vacuum with quasi-harmonic oscillations.  This feature in turn pave the way for the studying phase transition and matter generation in the bounce universe context~\cite{matter, baryon}.

A detailed analysis of the cosmological perturbations
 of this model has been carried in~\cite{Li:2012vi, Li:2013bha}.
 The spectrum  of the primordial matter  perturbations  takes the form
  ({\it c.f } Eq.(3.18) in ~\cite{Li:2012vi, Li:2013bha}):
\begin{equation}
P_\zeta=\left(\frac{H}{\dot{T}_c}\right)^2P_{\delta T}=\frac{C^2}{16\pi\kappa}k^0\eta^0
\end{equation}
with  $k$ being the wave vector and $\eta$ being the conformal time.
$P_{\delta T}$ is the spectrum of tachyon field whose perturbations
are responsible for primordial matter generation~\cite{Li:2012vi, Li:2013bha}.
Both $C$ and $\kappa$ are constant parameters whose  values are determined by
$\lambda$ and $\frac{M_s}{M_p}$.
Therefore, the primordial curvature perturbation of this model is scale invariant and stable.
Furthermore, taking the sub-leading effects into account, the spectral index of
the primordial curvature perturbation is also obtained in Eq.(3.22) 
of~\cite{Li:2012vi, Li:2013bha}. It is thus  natural obtain the  value of $n_s-1\le 0.04$
around a few percents  in  the CSTB model.
A detailed analysis presented in~\cite{Li:2012vi, Li:2013bha} shows  that
 a large parameter space in CSTB model satisfies the  current observational constraints of spectral index.

\subsection{How to make a successful bounce: Initial conditions}

In this section we study the constraints on parameters  from having enough
e-foldings.
In the locked inflation era  the universe is dominated by  vacuum energy.
The Hubble parameter is nearly constant as long as the tachyon is locked
at its peak by the scalar field $\phi$.
The dynamic equations of the Hubble parameter, $\phi$ and $T$
can hence be simplified as follows,
\begin{eqnarray}\label{simpH}
H^2 - \frac{8\pi{V_0}}{3M_p^2} &=&0 \\
\label{simpphi}
\ddot{\phi}+3H\dot{\phi}+m_\phi^2\phi&=&0 \\
\label{simpT}
\frac{1}{M_s^4}\ddot{T}+\left(-\frac{1}{2M_s^2}+\frac{2\lambda\phi^2}{V_0}\right)T &=& 0.
\end{eqnarray}
The scalar field, $\phi$, oscillates while being redshifted
\begin{equation}\label{phi}
\phi\propto{a^{-\frac{3}{2}}e^{i\sqrt{m^2_\phi-\frac{9}{4}H^2}t}}.
\end{equation}
We can calculate the number of  e-foldings  $N_L$ in this era
determined by the critical value of
$\langle{\phi_c^2}\rangle = {V_0/4\lambda{M_s^2}}$:
\begin{equation}\label{efolding}
   N_L=\frac{1}{6}
\ln{\frac{32\lambda^3M_s^4{\langle\phi_0^2\rangle}^2 \langle{T}_0^2\rangle}{m^2V_0^2}}.
\end{equation}

\begin{figure}[htbp]
\centering
\includegraphics[width=2in]{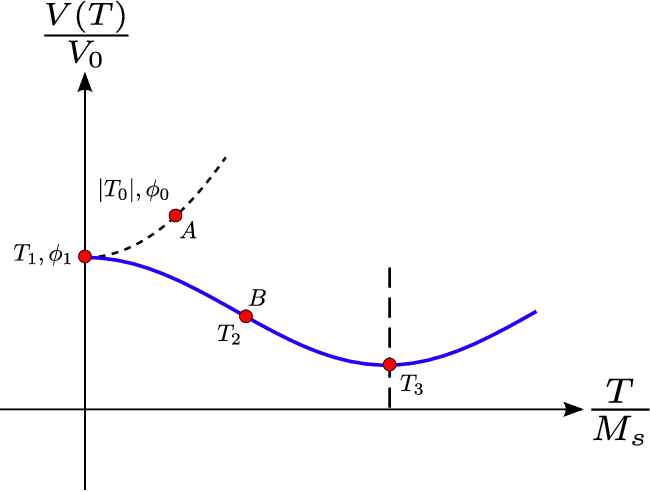}
\caption{A generic cross section ($\phi \ne 0$) of the tachyon-scalar field space: During the contraction period the tachyon field, $T$,  blue-shifts from $T_3$ to $T_2$  as the universe contracts from Point $C^\prime$ to Point $B^\prime$ in Fig.\ref{fig:bounce-model}.
When $T$ field goes to $T_1$, the effective mass of  $T$  becomes positive (dash line from $T_1$ to $T_0$), the universe is dominated by  vacuum energy and
 undergoes a deflation until the Hubble parameter becomes zero.
When $H=0$, the tachyon field is  locked at $T_0$, and the universe starts to inflate.}
\label{Tfield}
\end{figure}

Let us now turn to observational constraints on CSTB model:
\begin{itemize}
\item\emph{A period of vacuum energy domination:}
The first constraint comes from vacuum energy domination.
In the locked inflation era, the tension of D-$\mathrm{\bar{D}}$-brane
is so large that it dominates the energy density in this era.
\begin{equation}\label{cons1}
V_0>\frac{1}{2}m^2\langle\phi_0^2\rangle+\lambda\langle{T_0^2}\rangle\langle{\phi_0^2}\rangle
\end{equation}

\item\emph{The bounce point:}
For an extended period of locked inflation about  $T\sim 0$, the initial value
$\langle\phi_0^2\rangle$ should be larger than $\langle\phi_1^2\rangle$.
And the locked inflation ends when
\begin{equation}\label{cons2}
\langle\phi_0^2\rangle  < \frac{V_0}{4\lambda{M_s^2}}
\end{equation}

\item\emph{The effective  mass of $\phi$: }
The effective mass of $\phi$ is given by
$m_{eff}= m_{\phi}^2 + 2\lambda{T}^2$.
During locked inflation, $T$ red-shifts and its $vev$
$\langle{T^2}\rangle$ diminishes dramatically.
We assume that at the beginning of the locked inflation era $T_0$ is very large
and that the $T$ field is smaller than the bare mass of scalar field,
$m_{\phi}$  at the end. Therefore we have
\begin{equation}
\label{cons3}
2\lambda\langle{T_0^2}\rangle>m_{\phi}^2>2\lambda\langle{T_1^2}\rangle
\end{equation}
and  $\langle{T_1^2}\rangle$ can be expressed as $\sqrt{\frac{m_{\phi}^2V_0\langle{T_0^2}\rangle}{8\lambda^2M_s^2\langle{\phi_0^2}\rangle}}$.

\item \emph{Keeping $\phi$ oscillating:}
To keep $\phi$ oscillating in the locked inflation era, we demand the frequency of $\phi$ in~\ref{phi} be real, which in turn implies  $m_{\phi}^2\phi>\frac{9}{4}H^2$.
\begin{equation}\label{cons4}
2\lambda\langle{T_0^2}\rangle>\frac{6\pi}{M_p^2}(V_0+2\lambda\langle{{T}_0^2}\rangle\langle{\phi_0^2}\rangle).
\end{equation}
\end{itemize}
With these constraints we deduce  two critical relations amongst
the key parameters $m_{\phi}$, $M_s$ and $\lambda$:
\begin{eqnarray}
0<\frac{m}{\lambda{M_p}}<\frac{4(\sqrt{2}-1)}{\sqrt{3\pi}}
&\sim &0.54,  \\
\lambda &>& \frac{1}{8}.
\end{eqnarray}
We thus see that these values  cover a wide volume   of the parameter space,
and it is easy to pick  a generic set of  initial conditions for a successful
bounce model  resulting in a realistic universe. In particular this only
requires that  $N_L\sim 6$ e-foldings  are generated in the  locked inflation era.
This is to be contrasted with the fine tuning one needs for  successful inflation modelling.

\section{The contraction era in CSTB Universe}
We shall  show that CSTB universe is free of the Horizon problem
by calculating the e-foldings in the contraction phase of the CSTB universe.

\subsection{CSTB model solves the horizon problem}
\label{sec:horizon}
Bounce  universe models   solve the  horizon problem because
there exists  a phase of contraction--prior to the bounce--long enough to put the entire universe back into thermal contact and re-establish causality.
In CSTB model, particle horizon is therefore nonzero.
Considering that particle horizons in expansion and contraction phases are symmetric, we only calculate the particle horizon in expansion phase.

In the era of locked inflation, the universe is vacuum energy dominated:
 the Friedman equation becomes equation~\ref{simpH} and
 the  universe undergoes an exponential expansion
\begin{equation}
a=a_0 e^{\sqrt{\frac{8\pi V_0}{3M_p^2}}t}.
\end{equation}
The particle horizon in locked inflation is given by
\begin{equation}
d_{p_1}=a_1 \int^{a_1}_{a_0}\frac{da}{Ha^2}
=\frac{a_1}{H_0}(\frac{1}{a_0}-\frac{1}{a_{max}})~.
\end{equation}

Shortly after the bounce the Hubble parameter changes  from
$0$ to $\sqrt{\frac{8\pi V_0}{3M_p^2}}$,
and $H_0 a_0=1$, which in turn implies
$$d_{p_1}=a_1-a_0.$$
In the tachyon matter dominated era, the growth of the particle horizon is
 $$d_{p_2}=a_{max} \int^{a_{max}}_{a_1}\frac{da}{Ha^2}=\frac{2a_0 a_{max}}{a_1^{\frac{3}{2}}}(\sqrt{a_{max}}-\sqrt{a_1}).$$

All in all the total particle horizon at the end of expansion
would be
$$
d_{p_{tot}}=d_{p_1}+d_{p_2} =a_1-a_0+
\frac{2a_0 a_{max}}{a_1^{\frac{3}{2}}}
(\sqrt{a_{max}}-\sqrt{a_1});
$$
and thus $\frac{d_{p_{tot}}}{a_0}\gg1$.
Since the maximum size of universe is larger than current size $a_0$ at the bounce point the particle horizon is much larger than the scale factor.
So bounce   universe models  guarantee that the particle horizon be larger than the current homogeneous regions.

\subsection{E-foldings  of the universe}

In Big Bang cosmos the flatness problem is a cosmological fine-tuning problem.
CMB data~\cite{CMB} determines the current universe to be flat up to $\sim 1\%$ percent level, $\Omega=1.00\pm0.01$,
$$
\frac{1}{a_{now}^2H_{now}^2}<0.01. 
$$
As shown in Fig.~\ref{fig:bounce-model} we can read off  the required
e-foldings from the end of locked inflation to the present:
\begin{eqnarray}  \label{requiredefolding}
N_r =\ln{(\frac{a_1}{a_{now}})}
&=&\ln{(\frac{\rho_{now}}{\rho_D})^\frac{1}{3}(\frac{\rho_D}{\rho_C})^\frac{1}{4}(\frac{\rho_C}{\rho_B})^\frac{1}{3}}\\ \nonumber
&=&\ln{(\frac{T_{now}}{T_D})^\frac{4}{3}(\frac{T_D}{T_C})(\frac{T_C}{T_B})^\frac{4}{3}}~.
\end{eqnarray}

After the tachyon condenses the single tachyon field rolls towards infinity with almost zero effective mass.
When $\langle{T}\rangle$ becomes large, the tachyon field can be replaced by an equivalent scalar field
$\sigma=\frac{4\sqrt{V_0}}{M_s}e^{-\frac{T}{2\sqrt{2}M_s}}$,
which approaches zero when $T\rightarrow\infty$.
With the new field the action can be  expanded around $\sigma\sim 0$ and to first order in $\lambda$,
\begin{equation}\label{newaction}
L_{\phi-\sigma} = -\frac{1}{2}
\sigma^2-\frac{1}{2}\partial_{\mu}\sigma\partial^{\mu}\sigma-\frac{1}{2}\partial_{\mu}\phi\partial^{\mu}\phi-\frac{1}{2}m^2\phi^2-\lambda\phi^2\sigma^2,
\end{equation}
i.e. at point $C^\prime$ in Fig.~\ref{fig:bounce-model}, with
\begin{equation}
\label{neweomH}
(\frac{\dot{a}}{a})^2=-\frac{1}{a^2}+
\frac{8\pi}{3M_p^2}\left[
\frac{1}{2}\sigma^2+\frac{1}{2}m^2\phi^2-\lambda\sigma^2\phi^2+\frac{1}{2}\dot{\phi}^2+\frac{1}{2}\dot{\sigma}^2 \right].
\end{equation}
Scalar fields, $\sigma$ and $\phi$,  then take on similar equations of  motion:
\begin{eqnarray}
\label{neweomphi}
\ddot{\sigma}+3H\dot{\sigma}+\left[(\frac{M_s}{2})^2-2\lambda\phi^2\right]\sigma&=&0 \\
\label{neweomsigma}
\ddot{\phi}+3H\dot{\phi}+\left(m^2-2\lambda\sigma^2\right)\phi&=&0.
\end{eqnarray}

From~(\ref{neweomH}) we can deduce that at $C^\prime$, $\langle\sigma^2\rangle$ and $\langle\phi^2\rangle$ are at their minima, and divide
the energy density equally:
\begin{equation}\label{minphi}
\langle\phi^2\rangle_{min}=\frac{3M_p^2}{8\pi{m}^2a^2}.
\end{equation}
During the subsequent contraction phase, as the scale factor $a$ decreases,
$\langle\sigma^2\rangle$ and $\langle\phi^2\rangle$ blue-shift,
leading to a decrease in the  effective masses.
Considering that
${m^2_\sigma}=(\frac{m_s}{2})^2-2\lambda\phi^2$,
$m_\sigma^2$ decreases to zero and $\sigma$
 loses its validity and one should returns to  $T$.
 At $B^\prime$ in Figure~\ref{fig:bounce-model}, the contraction phase
 turns into deflation.
We denote the amplitude of $\phi$ at this point by:
\begin{equation}\label{endingcontraction}
\langle\phi^2\rangle_e=\frac{m_s^2}{8\lambda}.
\end{equation}
E-foldings can be computed accordingly,
$$
N_c=\frac{1}{3}\ln{\frac{\langle\phi^2\rangle_e}{\langle\phi^2\rangle_{min}}},
$$
because
$\langle\phi^2\rangle\propto{a}^3$.
Upon substituting the current size of universe
$a_{current}\sim 5.4\times10^{61}l_p$ into $N_c$, we push back  the
current universe to the beginning of the deflation era:
\begin{equation}\label{e-folding}
N_c=\frac{1}{3}\ln{\frac{8\pi{m}^2m_s^2a^2}{8\lambda3M_p^2}}\sim 118,
\end{equation}
where $m\sim\frac{1}{10}m_s \sim \frac{1}{100}M_p, \lambda \sim 0.25$.
It is persuasive to say that, with a contraction which can generate 120 e-foldings, the expansion era after the bounce can enjoy an analogous number of e-foldings~\cite{reviews}.
We can thus conclude  that CSTB cosmology  model  solves the  flatness problem by the 118 e-foldings  in the contraction phase.

\section{Conclusion and outlook}

We present a viable alternative  to inflation based on string theory utilising coupled scalar and tachyon fields in the D-$\mathrm{\bar{D}}$-brane system~\cite{CSTB}.
The CSTB cosmos undergoes a  period of tachyon matter dominated contraction
in which a stable and scale invariant spectrum of primordial density perturbations is  produced~\cite{Li:2012vi, Li:2013bha}.
The model has a signature out-of-thermal-equilibrium production of dark matter
which can be tested by the future array of dark matter
detections~\cite{Li:2014era, Cheung:2014nxi, Cheung:2014pea, Vergados:2016niz, Cheung:2016wik},  independently of cosmological observations~\cite{Cai:2014bea}.
Exploration on matter production and  baryon asymmetry  genesis is well underway~\cite{matter, baryon}.
We expect our  success can inspire further studies on bounce  models built from string theory or other quantum gravity theories.

Recently an interesting attempt is made to use AdS/CFT correspondence to study the evolution of matter through the bounce point~\cite{Ming:2017dtm} which is 
inspired by an earlier attempt of using AdS/CFT to study cosmic singularity in the bound universe~\cite{Brandenberger:2016egn}.  
Pioneer works on studying  AdS
singularity can be found in~\cite{Kumar:2015gln, Bzowski:2015clm, Kumar:2015jxa, Barbon:2015ria, Engelhardt:2015gla, Engelhardt:2015gta, Enciso:2015qva, Banerjee:2015fua}, 
together with earlier initiatives of using AdS/CFT to study cosmic singularities~\cite{Balasubramanian:1998de, Hertog:2005hu, Hamilton:2005ju, 
Chu:2006pa, Das:2006dz,Chu:2007um, Turok:2007ry, Awad:2007fj, Craps:2007ch, Barbon:2011ta, Enciso:2012wu, Engelhardt:2014mea}. 
  A thorough study of physics  in the bounce universe is in sight. 
   
Another interesting question to ask if the CST Bounce universe model can be
imbedded in higher dimensions, a la~\cite{Randall:1999vf}, and to study the
 stabilisation of the extra moduli in this context.
Further explorations shall also exploit the symmetries and dualities of the mother string theory  as  string-cosmology  models  utilising 
T-duality~\cite{Brandenberger:1988aj, Kounnas:2011fk}.  
This direction is expected to hold  more promise of modulus stabilisation~\cite{Berndsen:2004tj, Watson:2004aq, Brandenberger:2005bd}.

\Acknowledgements{We would like to thank Jin U Kang,  Changhong Li and 
Yuan Xin for useful discussions.
This research project has been supported in parts by the NSF China under Contract No.~11775110 and No.~11690034.
We also acknowledge the European Union's Horizon 2020 Research and Innovation~(RISE) programme under the Marie Sk\'lodowska-Curie grant 
agreement No.~644121,
and  the Priority Academic Program Development for
Jiangsu Higher Education Institutions (PAPD).}

\end{multicols}
\end{document}